\documentclass[11pt,a4paper]{article}
\usepackage[utf8]{inputenc}
\usepackage[T1]{fontenc}
\usepackage{lmodern}
\usepackage{microtype}
\usepackage[margin=2.5cm]{geometry}
\usepackage{setspace}
\usepackage{parskip}
\usepackage{wrapfig}
\usepackage{graphicx}
\usepackage{booktabs}
\usepackage{enumitem}
\usepackage{hyperref}
\hypersetup{colorlinks=true,linkcolor=blue,citecolor=blue,urlcolor=blue}
\usepackage{titling}
\usepackage{fancyhdr}
\usepackage{caption}
\usepackage{amsmath,amssymb}

\pretitle{\vspace*{-1cm}\begin{center}\LARGE\bfseries}
\posttitle{\par\end{center}\vskip 0.5em}
\preauthor{\begin{center}\large}
\postauthor{\end{center}}
\predate{\begin{center}\large}
\postdate{\end{center}}

\begin{document}

% -----------------------
% Cover page
% -----------------------
\begin{titlepage}
  \centering
  \vspace*{1cm}
  {\Huge\bfseries Expanding Horizons \\[6pt] \Large Transforming Astronomy in the 2040s \par}
  \vspace{1.5cm}

  % Title (editable)
  {\LARGE \textbf{The final stages of binary evolution using multi-messenger studies }\par}
  \vspace{1cm}

  % Metadata block
  \begin{tabular}{p{4.5cm}p{10cm}}
    \textbf{Scientific Categories:} & gravitational wave sources, multi-messenger astrophysics, binary evolution, compact objects \\
    \\
    \textbf{Submitting Author:} & Name: Thomas Kupfer \\
    & Affiliation: University of Hamburg \\
    & Email: thomas.kupfer@uni-hamburg.de \\
    \\
    \textbf{Contributing authors:} & Simone Scaringi (Durham University, UK)
Paul Groot (Radboud University, NL)
Boris Gänsicke (University of Warwick, UK)
Ingrid Pelisoli (University of Warwick, UK)
Anna F. Pala (ESO, DE)
Jan van Roestel (ISTA, AT; UvA, NL)
Silvia Toonen (UvA, NL)
Domitilla de Martino, (INAF-OACNa, IT)
Noel Castro Segura (University of Warwick, UK)
David Buckley (South African Astronomical Observatory, ZA)
Valerie Van Grootel (University of Liège, BE)
Kieran O’Brien (Durham University, UK)
Samaya Nissanke (DESY, DZA, University of Potsdam, DE; Amsterdam, NL)
 \\
  \end{tabular}

  \vspace{1cm}

  \textbf{Abstract:}

  \vspace{0.5em}
  \begin{minipage}{0.9\textwidth}
    \small
Ultracompact Galactic binaries with orbital periods below an hour are among the strongest persistent gravitational‑wave sources in the mHz band and will constitute the dominant population detected by the Laser Interferometer Space Antenna (LISA). Tens of thousands are predicted to be individually resolved, with a substantial fraction bright enough for electromagnetic (EM) follow‑up. This opens an unprecedented multi‑messenger window on compact binary evolution, tidal interactions, mass transfer, and the progenitors of Type Ia supernovae. We highlight key science enabled by joint GW+EM constraints and emphasize the critical need for rapid, high‑cadence spectroscopic capabilities in the 2040s. In particular, the most compact (<10 min) binaries detected by LISA will require read‑noise–free, zero‑dead‑time spectroscopic facilities—potentially realized through coordinated arrays of telescopes with time‑staggered exposures—to measure radial velocities, tidal heating signatures, and orbital evolution with the precision needed for transformative multi‑messenger studies.
  \end{minipage}

\end{titlepage}

% -----------------------
% Main sections (match DOCX template)
% -----------------------

\section*{Introduction}
Ultracompact binaries (UCBs) are a class of (semi-)detached binary stars with orbital periods as short as a few minutes and separations of the order of the Earth-Moon distance. This is only possible if both components are evolved objects consisting of a neutron star (NS), white dwarf (WD) or in rare cases black hole (BH) primary and a Helium-star/WD/NS secondary. These systems are the dominant Galactic sources in the LISA gravitational wave band, are crucial to our understanding of compact binary evolution, and offer pathways towards Type Ia supernovae. 

The formation mechanism for most UCBs involves at least one common envelope phase that produces a detached double WD binary system, which is brought into contact at a period between 3-10 minutes through gravitational wave emission (Podsiadlowski et al. 2003). In rare cases, if either one of the stars has an initial mass of M > 8-10 Msun the star will explode as a supernova after the first or second phase of mass transfer and end up as a NS or BH rather than a WD. This results in either a WD + NS/BH or a He-star + NS/BH system. If the UCB survives the onset of mass-transfer [this process and the stability of mass transfer are poorly understood, Nelemans et al. 2001; Marsh et al. 2004], and accretes stably onto a WD or a NS, the result is an AM CVn or an ultracompact X-ray binary (UCXB), respectively. Stable mass transfer in these systems leads to longer periods as the orbits widen. If the system does not survive the onset of mass-transfer, the two compact objects merge and either form an RCrB star or explode as a supernova Ia if the total mass is above the Chandrasekhar limit.

LISA is a space-based gravitational wave detector sensitive to lower frequencies than LIGO. UCBs are strong gravitational wave (GW) sources and will dominate the population of gravitational wave emitters in the LISA band. Systems with orbital periods <20min will be the strongest Galactic LISA sources and will be detected by LISA within weeks after its operation begins. These ‘verification binaries’ are crucial in facilitating the functional tests of the instrument and maximizing LISA’s scientific output. Galactic binaries with periods around 2 hrs are predicted to be so numerous that individual detections are limited by confusion with other binaries yielding a stochastic foreground or confusion signal. 

Binary population studies predict that we will be able to detect in GW and electromagnetic (EM) several thousand detached and semi-detached double WDs as well as a few tens of NS or black hole binaries with a population strongly peaking towards the Galactic Plane/Bulge (e.g. Nelemans et al. 2004, Lamberts et al. 2019). Observationally, we know of only about three dozen such systems, although hundreds to thousands are predicted to be detectable in our Galaxy (Fig. 1; Kupfer et al. 2024). Additionally, the known verification binary sample is strongly biased and incomplete, with most known sources located in the Northern hemisphere. Here, we present an overview of potential science for Galactic binaries using multi-messenger observations and finish with an ideal observational facility through the ESO Horizons 2040 program.

\section*{Astrophysics using combined EM+GW observations}
WD binaries will be among the best laboratories for understanding the formation of compact objects, dynamical interactions and tides in binaries, and Type Ia supernovae. Combined EM/GW observations yield more robust masses, radii, orbital separations, and inclination angles than either can achieve on its own. Shah, van der Sluys, \& Nelemans (2012) found a strong correlation between GW amplitude and inclination, and demonstrated that EM constraints on the inclination can improve the GW amplitude measurements by up to a factor of six. In addition, knowing the sky position and inclination can reduce the uncertainty in amplitude by up to a factor of 60 (Shah, Nelemans, \& van der Sluys 2013). \begin{wrapfigure}{r}{0.6\textwidth}
\vspace{-10pt}
  \centering
  \begin{minipage}{0.6\textwidth}
    \centering
    \includegraphics[width=\linewidth]{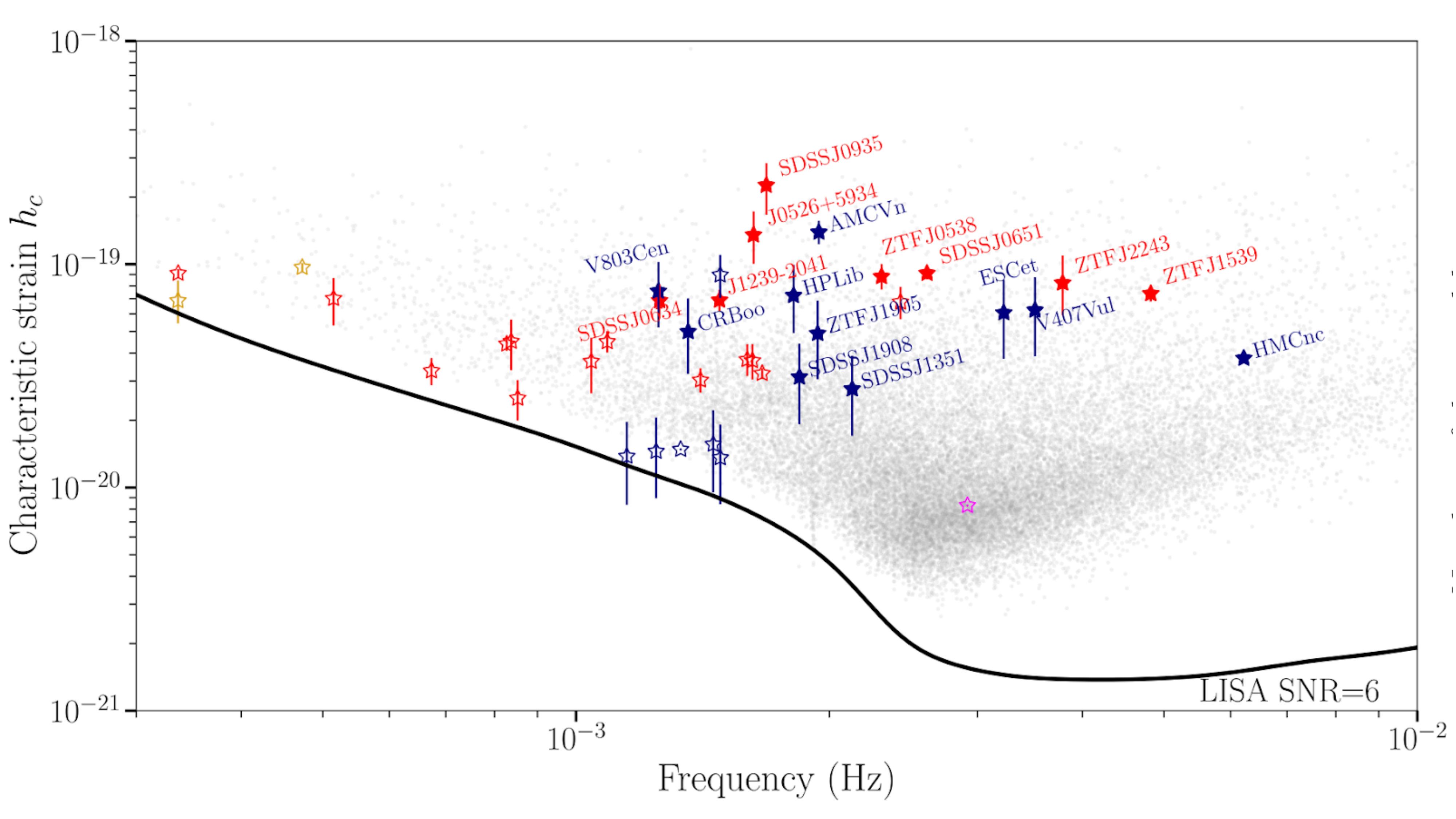}
  \end{minipage}\hfill
\caption{\small Characteristic strain–frequency plot for detectable and verification binaries: AM CVns in blue, DWDs in red, sdBs in yellow, and UCXB in magenta. Filled stars represent binaries detectable within 3 months of observations, which here we call verification binaries. The black solid line represents LISA's sensitivity curve that accounts for the instrumental noise (LISA Science Study Team 2018) and Galactic confusion foreground (Babak et al. 2017). For comparison in gray, we show a mock Galactic DWD population detectable with LISA from Wilhelm et al. (2021). Taken from Kupfer et al. (2024)}
\label{figure1}
\vspace{-15pt}
\end{wrapfigure} Similarly, using the chirp mass obtained from GW observations and the mass ratio from spectroscopic radial velocity measurements allows an independent measurement of the masses of both components to exquisite precision. This enables a direct comparison between the rate of orbital decay observed in GW or EM (for eclipsing and/or tidally distorted systems) and predicted from GR. 

Measuring the effects of tides in binaries, which are predicted to contribute up to 10\% of  the orbital decay is almost impossible from GW data alone (Shah \& Nelemans 2014), but the EM data on distance constrains the uncertainty in chirp mass to 20\%, whereas adding a measured orbital decay reduces it to 0.1\%. A GW chirp mass measurement would provide the first detection of tidal heating in a merging pair of WDs from the deviations in predicted orbital decay from GW alone. 
Prospects of Multi-messenger observations of Galactic binaries in the 2040s

Binary population studies predict that LISA will resolve several tens of thousands of Galactic ultracompact binaries (e.g. Nelemans et al. 2004; Korol et al. 2017). Systems with orbital periods of only a few minutes will be detectable across the entire Milky Way, with many entering the LISA catalog within the first weeks of operation due to their large gravitational-wave amplitudes and strong frequency evolution. Their high signal-to-noise ratios will allow precise localization and distance measurements, making them prime targets for coordinated multi-messenger observations.

The currently known population of verification binaries is strongly biased and heterogeneous, largely shaped by targeted searches and northern-hemisphere selection effects. This situation is expected to improve substantially in the coming decade. Ultracompact binaries can be detected through time-series photometry, which captures variability on timescales comparable to the orbital period (e.g. eclipses, ellipsoidal modulation, Doppler beaming), as well as through multi-epoch spectroscopy, where large radial-velocity shifts reveal tight orbits. As a result, ongoing and upcoming large-scale photometric and spectroscopic surveys—BlackGEM, ZTF, VRO (LSST), Gaia, 4MOST, SDSS-V, DESI, among others—are well positioned to identify LISA sources in a homogeneous and systematic manner. These will be complemented by continued surveys across the electromagnetic spectrum, from radio to gamma rays.

After four years of operation, LISA is expected to provide individual detections for tens of thousands of binaries, including several thousand (a few hundred with G<24mag) with sky-localization uncertainties smaller than 1 deg² and distance errors below 10\% (e.g. Littenberg and Cornish 2023). This population will form an unparalleled dataset for Galactic binary population studies and will offer a rich sample of targets for detailed EM follow-up.

A large number of these systems—especially the earliest verification binaries—will enable high-precision multi-messenger constraints on astrophysical phenomena central to compact binary evolution. These include the progenitors of thermonuclear supernovae, the stability of mass transfer, tidal heating and dissipation, and accretion physics under extreme conditions (see Amareo-Seoane et al. 2023 for a recent review). The most compact binaries, with orbital periods below 10 minutes, are of particular scientific interest: they are expected to exhibit the strongest tidal interactions with high accretion rates, the most rapid orbital evolution, and—in some cases—are within centuries of merging. Obtaining measurements of masses, radii, inclinations, accretion rates and orbital evolution for these sources requires rapid, continuous, high-cadence spectroscopy that is not achievable with any existing or currently planned single-aperture facility.

\section*{Technology and data handling requirements}
The most compact LISA binaries demand spectroscopy at cadences matched to their minute‑scale orbital periods. Requirements include:
\begin{itemize}
\item Zero readout noise to preserve high S/N at very short exposures.
 \item No dead time between exposures, essential for tracking rapid radial‑velocity variations.
\item Flexible, parallelizable operations to follow tens to hundreds or sources across the sky 
\end{itemize}
In particular, even a flagship observatory such as the ELT cannot provide the uninterrupted, dead-time–free, minute-cadence phase-resolved spectroscopy demanded by these ultracompact binaries. The limitations arise from finite readout times, instrument overheads, monolithic scheduling, and the practical impossibility of dedicating such a large facility to high-cadence monitoring of numerous faint sources. 

Instead, a promising concept could be an array of coordinated telescopes equipped with fast, low‑noise, and low to medium resolution spectrographs. By staggering exposure start times across the array, the system effectively achieves continuous coverage with no dead time, while each aperture individually avoids readout penalties. Such a facility would:
\begin{itemize}
    \item enable phase‑resolved spectroscopy for binaries with P < 10 min;
\item measure tidal heating, mass transfer rates, and orbital evolution;
\item provide EM anchors for GW localization, inclination, and chirp mass;
\item support long‑baseline monitoring of sources nearing contact or merger.
\end{itemize}
This capability is essential for extracting the full scientific potential of LISA’s most compact binaries and aligns naturally with ESO Horizons 2040 goals.

\section*{References}
Amareo-Seoane, P., et al. 2023, Living Reviews in Relativity, 26, 1 - Babak, S., et al. 2017, PhRv, D95, 103012 -  Korol, V., et al. 2017, MNRAS, 470, 1894 - Kupfer, T., et al. 2024, ApJ, 963, 2 - Lamberts, A., et al. 2019, MNRAS, 490, 4 - LISA Science Study Team 2018, LISA Science Requirements Document, Tech. Rep. ESA-L3-EST-SCI-RS-001, - Littenberg, T. B., \& Cornish, N. J. 2023, PhRvD, 107, 063004 - Marsh, T. R., et al. 2004, MNRAS, 350, 113 - Nelemans G., Yungelson L. R., Portegies Zwart S. F. 2004, MNRAS, 349, 181 - Nelemans, G., et al. 2001, A\&A, 368, 939 - Podsiadlowski, P., et al. 2003, MNRAS, 340, 1214 - Shah S., van der Sluys M., Nelemans G. 2012, A\&A, 544, A153 - Shah S., Nelemans G., van der Sluys M. 2013, A\&A, 553, A82 - Shah S., Nelemans G. 2014, ApJ, 791, 76 - Wilhelm, M. J. C., Korol, V., Rossi, E. M., \& D’Onghia, E. 2021, MNRAS, 500, 4958

\end{document}